\documentclass[a4,12pt]{article}
\usepackage{graphicx,psfrag,amssymb,cite}

\renewcommand{\thefootnote}{\fnsymbol{footnote}}

\newcommand{\prepr}[1] {\begin{flushright}  {\bf #1} \end{flushright} \vskip 1.cm}
\newcommand{\titul}[1] {\boldmath \begin{center}{\Large {\bf #1 } } \end{center}
\vskip 0.8cm}

\newcommand{\autor}[1] {\begin{center}  {\bf \lineskip .3cm #1  }
                        \end{center} }

\newcommand{\lugar}[1] {\begin{center}  {\normalsize \bf \it #1   } \end{center}}
\newcounter{muni}

\makeatletter
\def\fmslash{\@ifnextchar[{\fmsl@sh}{\fmsl@sh[0mu]}}
\def\fmsl@sh[#1]#2{%
  \mathchoice
    {\@fmsl@sh\displaystyle{#1}{#2}}%
    {\@fmsl@sh\textstyle{#1}{#2}}%
    {\@fmsl@sh\scriptstyle{#1}{#2}}%
    {\@fmsl@sh\scriptscriptstyle{#1}{#2}}}
\def\@fmsl@sh#1#2#3{\m@th\ooalign{$\hfil#1\mkern#2/\hfil$\crcr$#1#3$}}
\makeatother
\begin{document}
\hbadness=10000
\pagenumbering{arabic}
\begin{titlepage}

\prepr{hep-ph/0306037\\
\hspace{30mm} KIAS--P03034 \\
\hspace{30mm} TUM-HEP-514/03\\
\hspace{30mm} June 2003}

\begin{center}
\titul{\bf The effect of $H^\pm$ on $B^\pm\to \tau^\pm\nu_{\tau}$ and 
$B^\pm\to \mu^\pm\nu_{\mu}$
}

\autor{A.G. Akeroyd$^{\mbox{1}}$\footnote{akeroyd@kias.re.kr},
S. Recksiegel$^{\mbox{2}}$\footnote{stefan.recksiegel@ph.tum.de}}
\lugar{ $^{1}$ Korea Institute for Advanced Study,
207-43 Cheongryangri 2-dong,\\ Dongdaemun-gu,
Seoul 130-722, Republic of Korea}

\lugar{ $^{2}$ Physik-Department T31, 
Technische Universit\"at M\"unchen,\\
 D-85747 Garching, Germany }

\end{center}

\vskip2.0cm

\begin{abstract}
\noindent{The hitherto unobserved purely leptonic decays
$B^\pm\to \tau^\pm\nu_{\tau}$ and $B^\pm\to \mu^\pm\nu_{\mu}$ 
are of much interest at current and future runs of the 
$e^+e^-$ $B$ factories. Such decays are sensitive to charged
Higgs bosons ($H^\pm$) 
at the tree--level and provide important 
constraints on $\tan\beta/m_{H^\pm}$. We include the large corrections to the
$H^\pm ub$ coupling induced by virtual SUSY effects and show 
that the bounds on  $\tan\beta/m_{H^\pm}$ can 
be significantly weakened or strengthened.
}

\end{abstract}

\end{titlepage}
\thispagestyle{empty}
\newpage

\pagestyle{plain}
\renewcommand{\thefootnote}{\arabic{footnote} }
\setcounter{footnote}{0}

\section{Introduction}

The purely leptonic decays of the charged $B$ mesons, 
$B^\pm\to \tau^\pm\nu_{\tau}$ and $B^\pm\to \mu^\pm\nu_{\mu}$, 
are of much interest,
both in the Standard Model (SM) and in the context of models 
beyond the SM. The vast majority of $B^\pm$ mesons
(i.e. a bound state of 
$b\overline u$ or $\overline b u$) decay 
via the subprocess $b\to qX$, where $q=c,u,s,d$ (the
latter two via penguin operators) and $X$ are either
quarks or leptons.

\begin{figure}
\begin{center}
\includegraphics[width=6cm]{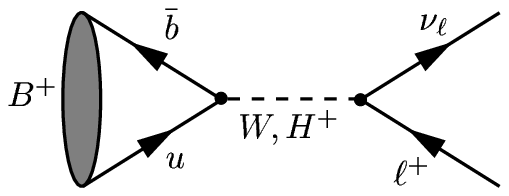}
\end{center}
\vspace*{-.5cm}
\caption{The decay $B^\pm\to \ell^\pm\nu_{\ell}$ is mediated
by annihilation into a virtual $W$ (and in extensions of the SM
also by a charged Higgs). \label{leptonicdecay}}
\end{figure}

A purely leptonic final state 
can occur via the annihilation process $B^\pm\to W^*\to l^\pm\nu_l$, 
although with a very suppressed rate (fig. \ref{leptonicdecay}).
Annihilation to $\mu^\pm\nu_{\mu}$ is the dominant decay
mechanism for the lighter charged mesons $\pi^\pm$ and $K^\pm$, 
while for $D^\pm_s$ the purely leptonic decays $\tau^\pm\nu_{\tau}$ 
and  $\mu^\pm\nu_{\mu}$
have been observed with subdominant branching ratios (BRs) 
in the range $10^{-2}\to 10^{-3}$ \cite{Hagiwara:fs}. 
No such purely leptonic decay has been observed for $B^\pm$ mesons.
Although their search is problematic at the hadronic B factories (e.g.\
Tevatron Run II) due to the missing energy of the $\nu$,
upper limits on their BRs were obtained at $e^+e^-$ colliders 
by CLEO \cite{Artuso:1995ar, Browder:2000qr}
and the LEP collaborations \cite{Acciarri:1996bv}. 
BELLE \cite{BELLE:2001} and BABAR \cite{Aubert:2003wu} 
are already producing more stringent limits, and expect 
to be sensitive to the SM rate by 2005/2006.
Observation of such decays would offer a direct measurement of 
the $B^\pm$ decay constant $f_B$.

These decays are also sensitive to many well motivated 
extensions of the SM \cite{Hou:1992sy}.
Charged Higgs bosons ($H^\pm$) appear in any model with two Higgs 
$SU(2)\times U(1)$ doublets (which
includes all supersymmetric models) and/or with Higgs 
triplets (which can provide neutrino mass) \cite{Gunion:1989we}.
In contrast to many other rare decays (e.g.
$b\to s\gamma$ ) which are influenced by new physics at the
one loop level, the decays $B^\pm\to l^\pm\nu_l$ are sensitive 
at tree--level to $H^\pm$ and can be enhanced up to the
current experimental limits \cite{Hou:1992sy}. 
From these upper limits
very useful constraints on the parameter $\tan\beta/m_{H^\pm}$
can be obtained (where $\tan\beta=v_2/v_1$, and $v_i$ are the 
vacuum expectation values of the Higgs doublets), 
which are comparable (and often superior) to those 
from processes like $B\to X\tau^\pm\nu$ \cite{Grossman:1994ax}
and $t\to H^\pm b$. The SM predictions for $B^\pm\to l^\pm\nu_{l}$
suffer from a sizeable uncertainty in $f_B$, whose
value can only be calculated using non--perturbative techniques
such as lattice QCD or inferred from other processes 
(e.g. $B^0\overline {B^0}$ mixing).
The current errors in such calculations 
induce an uncertainty in the SM BRs of around $30\%$ \cite{Ryan:2001ej}. 
It is likely that this sizeable error together with the small rates 
has dissuaded explicit calculations of higher order corrections, which
would na\"{\i}vely be expected to be at the $10\%$ level or less.
However, we turn out attention to $B^\pm\to \tau^\pm\nu_{\tau}$ and 
$B^\pm\to \mu^\pm\nu_{\mu}$ for three reasons.
\begin{itemize}

\item [{(i)}] Large corrections to the $H^\pm ub$ coupling
are possible in supersymmetric (SUSY) models for large $\tan\beta$
\cite{Hall:1993gn, Carena:1994bv, Blazek:1995nv}. 
Such corrections should significantly affect the BRs for 
$B^\pm\to \tau^\pm\nu_{\tau}$ and $B^\pm\to \mu^\pm\nu_{\mu}$ 
and the derived constraints on $\tan\beta/m_{H^\pm}$.

\item [{(ii)}] 
Due to the successful running of the $B$ factories,
the SM rate for both decays will be within experimental
sensitivity by 2005/2006. Higher Luminosity upgrades are now
being discussed \cite{unknown:2001xw}
which would offer an ideal environment for
precision studies of these decays.

\item [{(iii)}]
Improvements in lattice techniques resulting in more precise
predictions for $f_B$ \cite{Davies:2003ik}
are expected in the next few 
years. Data from the recently approved CLEO-c 
\cite{Shipsey:2002ye} will play an 
important role in refining the techniques.

\end{itemize}
The advances in (ii) and (iii) will facilitate 
the search for new physics in $B^\pm\to \tau^\pm\nu_{\tau}$ and 
$B^\pm\to \mu^\pm\nu_{\mu}$, and leads us to consider the 
higher order corrections to these decays in (i).

Our work is organised as follows. In Section 2 we
briefly review the tree--level contribution of $H^\pm$ and 
summarise the current experimental status.
In Section 3 we introduce the SUSY corrections while Section
4 presents our numerical results. Finally, Section 5
contains our conclusions.

\boldmath
\section{$B^\pm\to \tau^\pm\nu_{\tau}$ and $B^\pm\to \mu^\pm\nu_{\mu}$}
\label{Btotaunu}
\unboldmath

\begin{table}\begin{center}
\begin{tabular} {|c|c|c|c|c|} \hline
Decay & SM Prediction & CLEO   & {\sc Belle/\sc BABAR} & LEP \\ \hline
 $B^+\to e^+\nu_e$ & $9.2\times 10^{-12}$ & $\le 1.5\times 10^{-5}$
 \cite{Artuso:1995ar}& 
 $\le 5.4\times 10^{-6}$\cite{BELLE:2001} & $\otimes$ \\ \hline
 $B^+\to \mu^+\nu_{\mu}$ & $3.9\times 10^{-7}$ & 
$\le 2.1\times 10^{-5}$ \cite{Artuso:1995ar}
  & $\le 6.8\times 10^{-6}$ \cite{BELLE:2001} & $\otimes$ \\ \hline
  $B^+\to \tau^+\nu_{\tau}$ & $8.7\times 10^{-5}$ 
& $\le 8.4\times 10^{-4}$ \cite{Browder:2000qr} & 
$\le 4.1\times 10^{-4}$ \cite{Aubert:2003wu} 
& $\le 5.7\times 10^{-4}$ \cite{Acciarri:1996bv}  \\ \hline
\end{tabular}\end{center}
\caption{SM predictions and current experimental limits from various 
machines. \label{explimits}}
\end{table}

The $W^\pm$ and $H^\pm$ induce effective
four--fermion interactions \cite{Hou:1992sy} for 
$b\overline u\to l^-\overline\nu_l$ and $\overline b u\to l^+\nu_l$.
Below we write the Lagrangian for $b\overline u\to l^-\overline\nu_l$ induced
by $W^-$ and $H^-$ exchange:
\begin{equation}
{G_F\over \sqrt 2}V_{ub}\left(\left[\overline u\gamma_\mu(1-\gamma_5)b\right]\left[\overline l
\gamma^\mu(1-\gamma_5)\nu\right]-R_l\left[\overline u(1+\gamma_5)b\right]\left[\overline l
(1-\gamma_5)\nu\right]\right)
\end{equation}
with 
\begin{equation}
R_l=\tan^2\beta \,(m_bm_l/m_{H^\pm}^2)    \label{Rl}
\end{equation}
Here we only keep the dominant $m_b\tan\beta$ part of the $H^\pm ub$ 
coupling. The two quarks need to be at zero separation in order
to annihilate and this probability is contained in
the decay constant which is defined as:

\begin{equation}
<0|\overline u\gamma^\mu\gamma^5b|B^->=if_B p^\mu_B; \;\;\;\;\;
<0|\overline u\gamma^5b|B^->=-if_B (m_B^2/m_b)
\end{equation}
The decays proceed via the axial--vector part of the $W^\pm$
coupling and via the pseudoscalar
part of the $H^\pm$ coupling. Other couplings give a vanishing
contribution. The decay amplitude is therefore:

\begin{equation}
{\cal M}=-{G_F\over \sqrt 2}V_{ub}f_B\left[m_l-R_l(m^2_B/m_b)\right] \, \overline l
(1-\gamma_5)\nu
\end{equation}
\noindent
The tree--level partial width is given by \cite{Hou:1992sy}:
\begin{equation}
\Gamma(B^+\to \ell^+\nu_\ell)={G_F^2 m_{B} m_l^2 f_{B}^2\over 8\pi}
|V_{ub}|^2 \left(1-{m_l^2\over m^2_{B}}\right)^2 \,\times\, r_H
\end{equation}
where
\begin{equation}
r_H=[1-\tan^2\beta(m^2_B/m^2_{H^\pm})]^2
\end{equation}
The overall factor of $m_l^2$ arises from the helicity
suppression of the $W^\pm$ contribution, and the Yukawa coupling
of $H^\pm l^\pm\nu_l$.\footnote{Non--$m_l$--suppressed purely leptonic
decays are possible in other models, e.g.\ $R$--parity violating scenarios
\cite{RpV-Baek/AR}}
The dependence on $m_b$ cancels out in the 
tree--level approximation. This is because
$m_b$ in $R_l$ originates from the $b$ quark Yukawa coupling ($y_b$)
which at lowest order is simply related to $m_b$ by $y_b=\sqrt 2 m_b/v_1$.
Therefore $m_b$ in the numerator in ${\cal M}$
cancels with $m_b$ in the denominator.
At the 1-loop level the simple relationship
between $y_b$ and the running quark mass 
$m_b$ is modified, resulting in large corrections which will
be discussed in Section 3. 
One can see from the expression for $r_H$ 
that the $H^\pm$ contribution interferes destructively with 
that of $W^\pm$. Enhancement of these decays ($r_H>1$)
requires $\tan\beta/m_{H^\pm}> 0.27\,{\rm GeV}^{-1}$, 
while exact cancellation ($r_H=0$) of the $W^\pm$ and $H^\pm$ 
contributions occurs for $\tan\beta/m_{H^\pm}\approx 0.19\,
{\rm GeV}^{-1}$ ($=1/M_B$).

Table~\ref{explimits} displays the SM predictions for 
BR($B^\pm\to l^\pm\nu_{l}$) along with the current experimental limits 
from various machines.
We take $f_B=200$ MeV and $|V_{ub}|=0.0035$. 
The BRs are hierarchical due to the aforementioned $m_l^2$ suppression.
BR($B^\pm\to \tau^\pm\nu_{\tau}$) is much larger than 
BR($B^\pm\to \mu^\pm\nu_{\mu}$), but both decays are within
range of data samples with ${\cal O} (10^8)$ $B^\pm$ mesons. 
BR($B^\pm\to e^\pm\nu_e$) is too
small for any current or planned experiment.
The detection efficiency for $B^\pm\to \mu^\pm\nu_{\mu}$ 
is far superior to that for $B^\pm\to \tau^\pm\nu_{\tau}$,
which compensates for the smaller BR of the muonic channel. Hence
these two decays are competitive with each other for
probing new physics and/or for offering measurements of the decay
constant $f_B$. The upper limits in Table~1 on 
$B^\pm\to \tau^\pm\nu_{\tau}$ (from BABAR) and $B^\pm\to \mu^\pm\nu_{\mu}$ 
(from BELLE) constrain $r_H< 4.7$ and $17.4$ respectively.
For $f_B=200$ MeV this gives a bound on 
$\tan\beta/m_H<0.34\pm0.02$ GeV$^{-1}$ from $B^\pm\to \tau^\pm\nu_{\tau}$, 
with weaker limits from $B^\pm\to \mu^\pm\nu_{\mu}$. 
Data samples of order $4\times 10^8$ $B^\pm$ mesons, 
which are expected by 2006, should allow sensitivity
to the SM rate ($r_H=1$). BABAR simulations give yields
of 17 and 8 events (after all cuts) for the $\tau^\pm\nu_{\tau}$ and 
$\mu^\pm\nu_{\mu}$ channels respectively \cite{unknown:2001xw},
which would be enough to make
a first measurement of the BRs, or, in the case of
non-observation, would restrict $\tan\beta/m_{H^\pm}< 0.27$ GeV$^{-1}$.

\section{Higher order corrections}

Although the tree--level expression contains 
a sizeable error from the uncertainty in $f_B\,V_{ub}$ we feel that
including higher order corrections to the rates for 
$B^\pm\to \tau^\pm\nu_{\tau}$ and $B^\pm\to \mu^\pm\nu_{\mu}$
is well motivated, since current
experiments will reach the sensitivity of the SM prediction in the next
few years. In addition higher luminosity upgrades of both
PEP-II and KEKB are being discussed, which would provide data
samples of $10^9\!\to\! 10^{10}$ $B^\pm$ mesons and allow
precision measurements of these decays \cite{unknown:2001xw}.
Importantly, the $H^\pm$ contribution can already saturate the
current experimental limits (as shown in Section \ref{Btotaunu}), 
and so one or both of $\tau^\pm\nu_{\tau}$ and 
$\mu^\pm\nu_{\mu}$ might be observed in the near future.
The constraints on $\tan\beta/m_{H^\pm}$ derived from these
decays are among the most model independent 
constraints on this important ratio. We stress that other rare decays
(e.g. $B\to X_s\gamma, B_{d,s}\to \mu^+\mu^-$) can also give strong
constraints on $\tan\beta/m_{H^\pm}$
\cite{Bobeth:2001sq,Logan:2000iv,Chankowski:2000ng,D'Ambrosio:2002ex}. 

The observed inclusive decay 
$B^\pm\to X_c\tau^\pm\nu$ \cite{Barate:2000rc}
is also sensitive to $H^\pm$ at tree-level
and provides competitive, mostly model independent limits of 
$\tan\beta/m_{H^\pm}<0.5$. The exclusive decay $B^\pm\to D^0\tau^\pm\nu$
will be pursued at the $B$ factories \cite{Tanaka:1994ay}.
Constraints on $\tan\beta/m_{H^\pm}$ from the unobserved top 
quark decay $t\to H^\pm b$ at the Tevatron Run I are inferior to
those from $B^\pm\to \tau^\pm\nu$ and $B^\pm\to X_c\tau^\pm\nu$, but
will improve significantly with 2 fb$^{-1}$ expected during 
Run II \cite{Carena:2000yx}.
Given the importance of these constraints on
$\tan\beta/m_{H^\pm}$, we feel that higher order corrections to the
rates for BR($B^\pm\to \tau^\pm\nu$) and BR($B^\pm\to \mu^\pm\nu$) 
should be considered. We believe that a closer analysis of these
decays is timely due to the anticipated improvements 
in lattice calculations of $f_B$ in next few years \cite{Davies:2003ik}.

%
In this study we will only consider the sizeable corrections to the
coupling $H^\pm u_Id_J$ which occur at large $\tan\beta$ in SUSY models. 
The decays of interest to us depend on the coupling $H^\pm ub$ ($I=1,J=3$)
and the corrections are comprised of the following \cite{Buras:2002vd}:
\begin{itemize}

\item [{(i)}] Enhanced corrections to $H^\pm ub$ vertex diagrams

\item [{(ii)}] Large corrections to the $b$ quark 
Yukawa coupling $y_b$ \cite{Hall:1993gn,Carena:1994bv} (known as the HRS effect)

\item [{(iii)}] Large corrections to $V_{ub}$ \cite{Blazek:1995nv}, where
 $V_{ub}$ in the MSSM Lagrangian differs from the experimentally
measured  $V^{eff}_{ub}$

\end{itemize}

%
The above corrections greatly affect other processes involving the
$H^\pm u_Id_J$ coupling such as
$b\to s\gamma$ \cite{Carena:2000uj} and $t\to H^\pm b$ \cite{Carena:1999py}
(see \cite{Logan:2000cz} for a review).
The effects of these corrections on $B^\pm\to X_c\tau^\pm\nu$
were covered in \cite{D'Ambrosio:2002ex} but
the decays $B^\pm\to \tau^\pm\nu_{\tau}$ and $B^\pm\to \mu^\pm\nu_{\mu}$ 
of interest to us were overlooked.
It has been shown \cite{Buras:2002vd} that the combined effect of 
these corrections can be neatly encoded in an effective 
Lagrangian for the $H^\pm u_J d_I$ coupling by multiplying
the usual Lagrangian by a factor $1/(1+\tilde\epsilon_J\tan\beta$).
For our case of the decay of a $B_u$, $J=1$ and the corresponding
$\tilde\epsilon_J$ is called $\tilde\epsilon_0$ in \cite{Buras:2002vd}.

This correction appears in $R_l$ (eq.\ref{Rl}) and modifies 
the scaling factor of the SM rate to:
\begin{equation}
r_H=\left(1-{\tan^2\beta\over 1+\tilde\epsilon_0\,\tan\beta}\,{m^2_B\over 
m^2_{H^\pm}}\right)^2
\end{equation}

Analogous corrections of type (i) and (ii) are also
present for the $H^\pm \tau^\pm\nu_{\tau}$ 
and $H^\pm \mu^\pm\nu_{\mu}$ couplings, but are known to be 
considerably smaller since there is no SUSY QCD contribution 
(only SUSY electroweak
contributions). Thus we will neglect them in the present study. 
These corrections will be considered
in a future work and are of more interest for
high luminosity runs of the $B$ factories with data samples of
$10^{10}$ $B^\pm$ mesons. Note that both $f_B$ and the large
corrections to the $H^\pm ub$ coupling cancel out in the 
ratio $R_{\mu\tau}$ defined by: 
\begin{equation}
R_{\mu\tau}={BR(B^\pm\to \tau^\pm\nu_{\tau})
\over BR(B^\pm\to \mu^\pm\nu_{\mu})}
\end{equation}
This quantity is predicted to be $0.8\,y^2_\tau/y^2_\mu$,
which in the SM is essentially equivalent to $0.8\,m^2_\tau/m^2_\mu$. 
However, $R_{\mu\tau}$ would be affected by the HRS corrections 
to $y_{\tau}$ and the vertex corrections to $H^\pm l^\pm\nu_{\l}$.

\section{Numerical Analysis}
In this section we discuss the numerical impact of the modification
of the $H^\pm ub$ vertex on the combined bounds on $\tan\beta$ and
$m_{H^\pm}$. We take the modification parameter $\tilde\epsilon_0$
as an input parameter, varying it in the interval $[-0.01,0.01]$.
This is the range found in \cite{Buras:2002vd} by explicit calculation
and a scan over reasonable values of the Minimum Flavour Violation
Model parameters.

%
\begin{figure}
\begin{center}
\psfrag{170MeV}{$f_B=170\,{\rm MeV}$}\psfrag{200MeV}{$f_B=200\,{\rm MeV}$}
\psfrag{230MeV}{$f_B=230\,{\rm MeV}$}
\psfrag{eps=0}{$\tilde\epsilon_0=0$}\psfrag{eps=.01}{$\tilde\epsilon_0=0.01$}
\psfrag{eps=-.01}{$\tilde\epsilon_0=-0.01$}
\psfrag{mH}{$m_{H^\pm}$}\psfrag{tanb}{$\tan\beta$}\psfrag{tau}{\Huge $\tau$}
\psfrag{excluded}{excluded}\psfrag{area}{area}
\includegraphics[width=13cm]{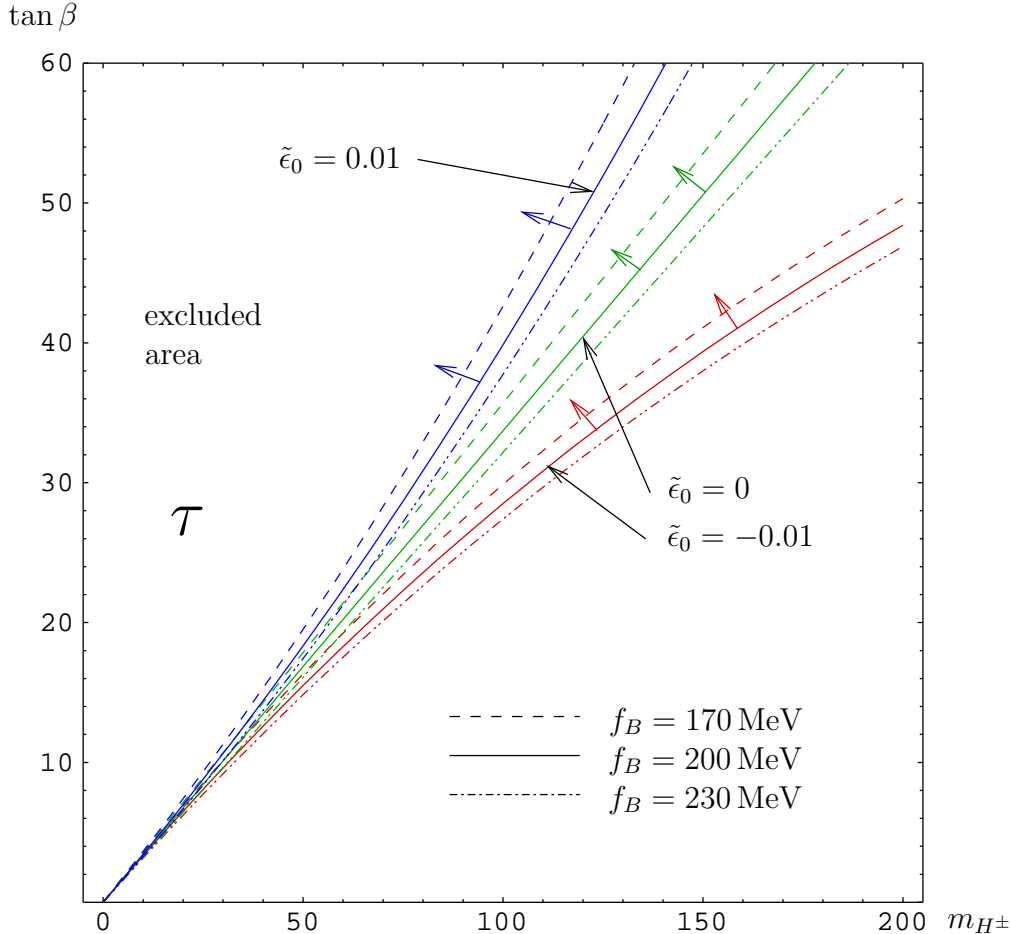}
\end{center}
\vspace*{-.5cm}
\caption{Excluded regions in the $\tan\beta-m_{H^\pm}$ plane for
different values of $\tilde\epsilon_0$ and $f_B$ from the experimental limit
${\rm BR}(B^+\to\tau\nu)<4.1 \cdot 10^{-4}$. The corresponding plot for
the $\mu$ case looks very similar but gives slightly weaker constraints.
\label{exclusionplot_tau}}
\end{figure}
In order to show the region in the $\tan\beta-m_{H^\pm}$ plane excluded by 
the experimental upper limits on BR$(B^\pm\to \tau^\pm\nu_{\tau})$,
in Fig.\ref{exclusionplot_tau} we 
plot $\tan\beta$ ($y$-axis) against $m_{H^\pm}$
($x$-axis). For a given set of input parameters, the region above the
corresponding curve is excluded. 

From BR$(B^\pm\to \tau^\pm\nu_{\tau})<4.1 \cdot 10^{-4}$,
without the radiative corrections, ($\tilde\epsilon_0=0$) the excluded region
is simply bounded by a straight line of gradient 
$R=0.34$ ($0.36, 0.32$) $[{\rm GeV}^{-1}]$ for $f_B=200$ ($170,230$) MeV. 
With the inclusion of radiative corrections at the level of
$\tilde\epsilon_0=-0.01$ (0.01), the bounds change to the lower (upper) set
of curved lines. It can be seen that even with the present rather large
uncertainty of $f_B$ of $\pm 15\%$, the effect of the radiative corrections
is much larger than that of the uncertainty of the decay constant.

With the central value of $f_B=200$ MeV, $m_{H^\pm}=120$ GeV would allow
$\tan \beta$ of up to 40 when the radiative corrections are neglected, 
strengthening to $\tan \beta < 33$ for $\tilde\epsilon_0=-0.01$ and 
weakening to
$\tan \beta < 49$ for $\tilde\epsilon_0=0.01$. The corresponding bounds from
BR$(B^\pm\to \mu^\pm\nu_{\mu})< 6.5 \cdot 10^{-6}$ are slightly weaker:
$\tan \beta < 51$ for $\tilde\epsilon_0=0$ and 40 (66) for 
$\tilde\epsilon_0=-0.01$ (0.01) (all for $f_B=200$ MeV, $m_{H^\pm}=120$ GeV).
Variations of $f_B$ have a much smaller effect. For the $\tau$ case
and with $m_{H^\pm}=120$ GeV, the bound on $\tan\beta$ changes from
40.5 (200 MeV) to 42.8 (38.6) for $f_B=170$ (230) MeV.

\section{Conclusions}
The unobserved purely leptonic decays $B^\pm\to \tau^\pm\nu_\tau$ 
and $B^\pm\to \mu^\pm\nu_\mu$ provide competitive constraints on
the ratio $\tan\beta/m_{H^\pm}$. We have studied the effect of SUSY 
corrections at large $\tan\beta$ and showed that the exact bounds 
on $\tan\beta/m_{H^\pm}$ can be strengthened or weakened depending on
the value of $\tilde\epsilon_0$, the correction factor for the
$H^\pm ub$ vertex due to SUSY loop corrections.
The effect on the allowed region in the ($\tan\beta, m_{H^\pm}$) plane
exceeds $\pm 20\%$ for $\tilde\epsilon_0=\pm0.01$, which is considerably
larger than the analogous uncertainty of $\pm 5\%$ caused by varying 
the decay constant $f_B$ by $\pm 30$ MeV.

\section*{Acknowledgements}
The authors wish to thank Abdesslam Arhrib for useful discussions.


\begin{thebibliography}{99}

\bibitem{Hagiwara:fs}
K.~Hagiwara {\it et al.}  [Particle Data Group Collaboration],
Phys.\ Rev.\ D {\bf 66}, 010001 (2002).

\bibitem{Artuso:1995ar}
M.~Artuso {\it et al.}  [CLEO Collaboration],
Phys.\ Rev.\ Lett.\  {\bf 75}, 785 (1995).

\bibitem{Browder:2000qr}
T.~E.~Browder {\it et al.}  [CLEO Collaboration],
Int.\ J.\ Mod.\ Phys.\ A {\bf 16S1B}, 636 (2001).

\bibitem{Acciarri:1996bv}
M.~Acciarri {\it et al.}  [L3 Collaboration],
Phys.\ Lett.\ B {\bf 396}, 327 (1997).

\bibitem{BELLE:2001}
BELLE--CONF--0127, http://belle.kek.jp/conferences/LP01-EPS/.

\bibitem{Aubert:2003wu}
B.~Aubert {\it et al.}  [BABAR Collaboration],
hep-ex/0303034.

\bibitem{Hou:1992sy}
W.~S.~Hou,
Phys.\ Rev.\ D {\bf 48}, 2342 (1993).

\bibitem{Gunion:1989we}
J.~F.~Gunion, H.~E.~Haber, G.~L.~Kane and S.~Dawson,
``The Higgs Hunter's Guide'', Addison-Wesley, 1990.

\bibitem{Grossman:1994ax}
Y.~Grossman and Z.~Ligeti,
Phys.\ Lett.\ B {\bf 332}, 373 (1994).

\bibitem{Ryan:2001ej}
S.~M.~Ryan,
Nucl.\ Phys.\ Proc.\ Suppl.\  {\bf 106}, 86 (2002).

\bibitem{RpV-Baek/AR}
S.~Baek and Y.~G.~Kim,
Phys.\ Rev.\ D {\bf 60}, 077701 (1999);
A.~G.~Akeroyd and S.~Recksiegel,
Phys.\ Lett.\ B {\bf 541}, 121 (2002).

\bibitem{Hall:1993gn}
L.~J.~Hall, R.~Rattazzi and U.~Sarid,
Phys.\ Rev.\ D {\bf 50}, 7048 (1994).

\bibitem{Carena:1994bv}
M.~Carena, M.~Olechowski, S.~Pokorski and C.~E.~Wagner,
Nucl.\ Phys.\ B {\bf 426}, 269 (1994);
R.~Hempfling,
Phys.\ Rev.\ D {\bf 49}, 6168 (1994).

\bibitem{Blazek:1995nv}
T.~Blazek, S.~Raby and S.~Pokorski,
Phys.\ Rev.\ D {\bf 52}, 4151 (1995).

\bibitem{unknown:2001xw}
``Physics at a 10**36 asymmetric B factory,''
SLAC-PUB-8970;
M.~Yamauchi,
Nucl.\ Phys.\ Proc.\ Suppl.\  {\bf 111}, 96 (2002).

\bibitem{Davies:2003ik}
C.~T.~Davies {\it et al.},
hep-lat/0304004.

\bibitem{Shipsey:2002ye}
I.~Shipsey,
hep-ex/0207091.


\bibitem{Bobeth:2001sq}
C.~Bobeth, T.~Ewerth, F.~Kruger and J.~Urban,
Phys.\ Rev.\ D {\bf 64}, 074014 (2001).

\bibitem{Logan:2000iv}
H.~E.~Logan and U.~Nierste,
Nucl.\ Phys.\ B {\bf 586}, 39 (2000).

\bibitem{Chankowski:2000ng}
P.~H.~Chankowski and L.~Slawianowska,
Phys.\ Rev.\ D {\bf 63}, 054012 (2001).


\bibitem{D'Ambrosio:2002ex}
G.~D'Ambrosio, G.~F.~Giudice, G.~Isidori and A.~Strumia,
Nucl.\ Phys.\ B {\bf 645}, 155 (2002).

\bibitem{Barate:2000rc}
R.~Barate {\it et al.}  [ALEPH Collaboration],
Eur.\ Phys.\ J.\ C {\bf 19}, 213 (2001).

\bibitem{Tanaka:1994ay}
M.~Tanaka,
Z.\ Phys.\ C {\bf 67}, 321 (1995);
K.~Kiers and A.~Soni,
Phys.\ Rev.\ D {\bf 56}, 5786 (1997);
T.~Miki, T.~Miura and M.~Tanaka,
hep-ph/0210051.

\bibitem{Carena:2000yx}
M.~Carena {\it et al.}  [Higgs Working Group Collaboration],
hep-ph/0010338.

\bibitem{Buras:2002vd}
A.~J.~Buras, P.~H.~Chankowski, J.~Rosiek and L.~Slawianowska,
hep-ph/0210145.


\bibitem{Carena:2000uj}
M.~Carena, D.~Garcia, U.~Nierste and C.~E.~Wagner,
Phys.\ Lett.\ B {\bf 499}, 141 (2001);
G.~Degrassi, P.~Gambino and G.~F.~Giudice,
JHEP {\bf 0012}, 009 (2000);
F.~Borzumati, C.~Greub and Y.~Yamada,
hep-ph/0305063.

\bibitem{Carena:1999py}
M.~Carena, D.~Garcia, U.~Nierste and C.~E.~Wagner,
Nucl.\ Phys.\ B {\bf 577}, 88 (2000);
J.~A.~Coarasa, D.~Garcia, J.~Guasch, R.~A.~Jimenez and J.~Sola,
Eur.\ Phys.\ J.\ C {\bf 2}, 373 (1998);
R.~A.~Jimenez and J.~Sola,
Phys.\ Lett.\ B {\bf 389}, 53 (1996)

\bibitem{Logan:2000cz}
H.~E.~Logan,
Nucl.\ Phys.\ Proc.\ Suppl.\  {\bf 101}, 279 (2001).

\end{thebibliography}
\end{document}